\documentclass[10pt,conference]{IEEEtran}
\usepackage{cite}
\usepackage{amsmath,amssymb,amsfonts}
\usepackage{algorithmic}
\usepackage{graphicx}
\usepackage{textcomp}
\usepackage{float}

\usepackage{xcolor}
\def\BibTeX{{\rm B\kern-.05em{\sc i\kern-.025em b}\kern-.08em
    T\kern-.1667em\lower.7ex\hbox{E}\kern-.125emX}}

\usepackage{multirow}
\usepackage[breaklinks=true]{hyperref}
\def\CC{{C\nolinebreak[4]\hspace{-.05em}\raisebox{.2ex}{++}}}

\usepackage{caption}

\begin{document}

\title{Coyote \CC: An Industrial-Strength Fully Automated Unit Testing Tool}

\author{
\IEEEauthorblockN{Sanghoon Rho}
\IEEEauthorblockA{\textit{Automated Testing Team} \\
\textit{CODEMIND Corporation}\\
Seoul, South Korea \\
rho@codemind.co.kr}\\
\IEEEauthorblockN{Yeoneo Kim}
\IEEEauthorblockA{\textit{Automated Testing Team} \\
\textit{CODEMIND Corporation}\\
Seoul, South Korea \\
yeoneo@codemind.co.kr}
\and
\IEEEauthorblockN{Philipp Martens}
\IEEEauthorblockA{\textit{Automated Testing Team} \\
\textit{CODEMIND Corporation}\\
Seoul, South Korea \\
philipp.m@codemind.co.kr}\\
\IEEEauthorblockN{Hoon Heo}
\IEEEauthorblockA{\textit{Control System Validation Team} \\
\textit{Hyundai KEFICO Corporation}\\
Gyeonggi-Do, South Korea \\
hoon.heo@hyundai-kefico.com}
\and
\IEEEauthorblockN{Seungcheol Shin}
\IEEEauthorblockA{\textit{Chief Executive Officer} \\
\textit{CODEMIND Corporation}\\
Seoul, South Korea \\
shin@codemind.co.kr}\\
\IEEEauthorblockN{SeungHyun Oh}
\IEEEauthorblockA{\textit{Control System Validation Team} \\
\textit{Hyundai KEFICO Corporation}\\
Gyeonggi-Do, South Korea \\
seunghyun.oh@hyundai-kefico.com}
}

\maketitle

\begin{abstract}
Coyote \CC\ is an automated testing tool that uses a sophisticated concolic-execution-based approach to realize fully automated unit testing for C and \CC. While concolic testing has proven effective for languages such as C and Java, tools have struggled to achieve a practical level of automation for \CC\ due to its many syntactical intricacies and overall complexity. Coyote \CC\ is the first automated testing tool to breach the barrier and bring automated unit testing for \CC\ to a practical level suitable for industrial adoption, consistently reaching around 90\% code coverage. Notably, this testing process requires no user involvement and performs test harness generation, test case generation and test execution with ``one-click'' automation. In this paper, we introduce Coyote \CC\ by outlining its high-level structure and discussing the core design decisions that shaped the implementation of its concolic execution engine. Finally, we demonstrate that Coyote \CC\ is capable of achieving high coverage results within a reasonable timespan by presenting the results from experiments on both open-source and industrial software.
\end{abstract}


\section{Introduction}

The significance of testing in software engineering is continuously escalating, necessitating thorough validation methods such as white-box testing. However, given the rapid increase in code scale and complexity in the software industry, white-box testing can be time-consuming and resource-intensive, often leading to budget constraints. For this reason, there has been a long-standing need for automation in white-box testing.

Lately, efforts to automate white-box unit testing are approaching practical feasibility, with automated testing showing promising results for Java~\cite{EvoSuite,jcute}, C~\cite{Maist,cute,dart}, binary code~\cite{sage, triton}, and a few programming languages~\cite{Pex,cuter,jalangi}. Conversely, adopting this technology for \CC\ has proven to be challenging due to the language's unique features and overall complexity. Implicitly invoked copy or move constructors and templates with all their intricacies are just two examples of \CC\ language features that are especially difficult to handle in automated white-box unit testing.

In this paper, we introduce Coyote \CC, an automated unit testing tool designed for C/\CC. With a single click, Coyote \CC\ streamlines the entire testing process, from harness generation and test case generation to test execution. The automated test case generation is based on concolic execution, a modern variant of symbolic execution, and features exquisite harness generation capabilities. 

The paper outlines the underlying technologies on which Coyote \CC\ achieves a practical level of high coverage through test case generation. In order to practically utilize automated unit testing tools in the field, we propose that a testing speed of around 10,000 logical LOC of executable statements per hour with statement coverage above 90\% and branch coverage above 80\% should be desirable. Currently, Coyote \CC\ is achieving elevated levels of coverage and performance according to these criteria, and is thus being effectively applied and utilized by our customers in the automotive industry.

The rest of this paper is organized as follows. We first look at research on concolic-execution-based unit testing and then examine design decisions made by existing systems to build efficient concolic execution engines in related works. Next, we provide an overview of the implementation of Coyote \CC, and present test results obtained from open-source projects and real-world industrial projects. Finally, we conclude the paper by outlining our plans for further improving Coyote \CC.

\section{Related works}

Symbolic execution~\cite{king1975} is a static program analysis technique that interprets programs with symbolic values rather than concrete values. Due to scalability issues with symbolic execution, this technique has been extended into concolic execution~\cite{cute, dart}. The main idea of concolic execution is to compute test inputs from path conditions which are obtained by tracking both concrete values and symbolic values. Concolic execution has been anticipated in the automated testing domain due to its known success in test case generation. However, this research has not yet reached a practical level of test generation for whole programs.

Nevertheless, concolic execution is known to be remarkably successful in unit test generation, e.g. for Java~\cite{EvoSuite,jcute} and C~\cite{Maist,cute,dart}. For \CC\ however, automated testing has still been far from viable for industrial purposes despite recent research efforts~\cite{Utbot,Citrus}.

When implementing concolic execution there are many options for realizing various aspects of the engine~\cite{SymExecSurvey}. Especially the engine's execution mode, analysis target, handling of the path explosion problem, and its memory model can largely affect the performance of the concolic execution engine in terms of coverage and execution time. \\[-6pt]

\subsubsection{Online/Offline Mode}
Concolic execution can be implemented in online or offline mode. In online mode, the concolic execution engine explores multiple paths in a single run by forking on branch points. The advantage of this method is that there is no need to re-execute the common prefixes of multiple paths. However, it requires a substantial amount of memory to store all the states of multiple paths. Offline mode on the other hand explores only one path in a single run. This method requires less memory than online mode, making it better suited for parallelization. However, since offline mode always starts at the beginning of the program for every path, it spends a considerable amount of time on re-examining common path prefixes. Prominent tools using online mode are KLEE~\cite{klee}, \textsc{Mayhem}~\cite{mayhem}, and S\textsuperscript{2}E~\cite{s2e}, whereas SAGE~\cite{sage} utilizes offline concolic execution. \\[-6pt]

\subsubsection{Emulation/Instrumentation}
There are two main methods for collecting information about the execution path taken during concrete execution of the program under test. The first method performs symbolic execution at the same time as concrete execution by running the program under test inside of an emulator such as QEMU~\cite{qemu}. The second method instead instruments the program under test with code that handles symbolic execution and the collection of information about the concrete execution of the program. Well-known emulator-based tools are angr~\cite{angr} and KLEE~\cite{klee}, while QSYM~\cite{qsym} and CREST~\cite{crest} are instrumentation-based. \\[-6pt]

\subsubsection{Mitigating Path Explosion}
Another important design decision is how to deal with the path explosion problem commonly encountered when performing concolic execution on programs with complex control flow. In such situations, the search space of concolic execution can grow exponentially due to the many possible combinations of branches. To avoid this issue, concolic execution engines use a variety of heuristic search strategies. Notable search strategies include DFS (depth-first search), BFS (breadth-first search), random path selection, coverage-optimized search, and adaptive heuristics~\cite{SymExecSurvey,searchHeuristics}. \\[-6pt]

\subsubsection{Memory Model}
When modelling the symbolic memory of a concolic execution engine, one can choose between treating memory addresses as symbolic or concrete values. The symbolic approach can theoretically handle all possible paths, but this approach may cause path constraints to become too complex for current SMT solvers. On the other hand, using concrete addresses might not cover all possible paths due to overly simplified path conditions. In practice, a fully symbolic model is used by tools like KLEE~\cite{klee}, and a concrete address model is used by SAGE~\cite{sage} among others. Additionally, there are tools like \textsc{Mayhem}~\cite{mayhem} that use a combination of symbolic and concrete addressing schemes.

\section[The Design of Coyote C++]{The Design of Coyote \CC}
\subsection[Overview]{Overview}
\begin{figure*}[!htbp]
    \centering
    \includegraphics[scale=0.7]{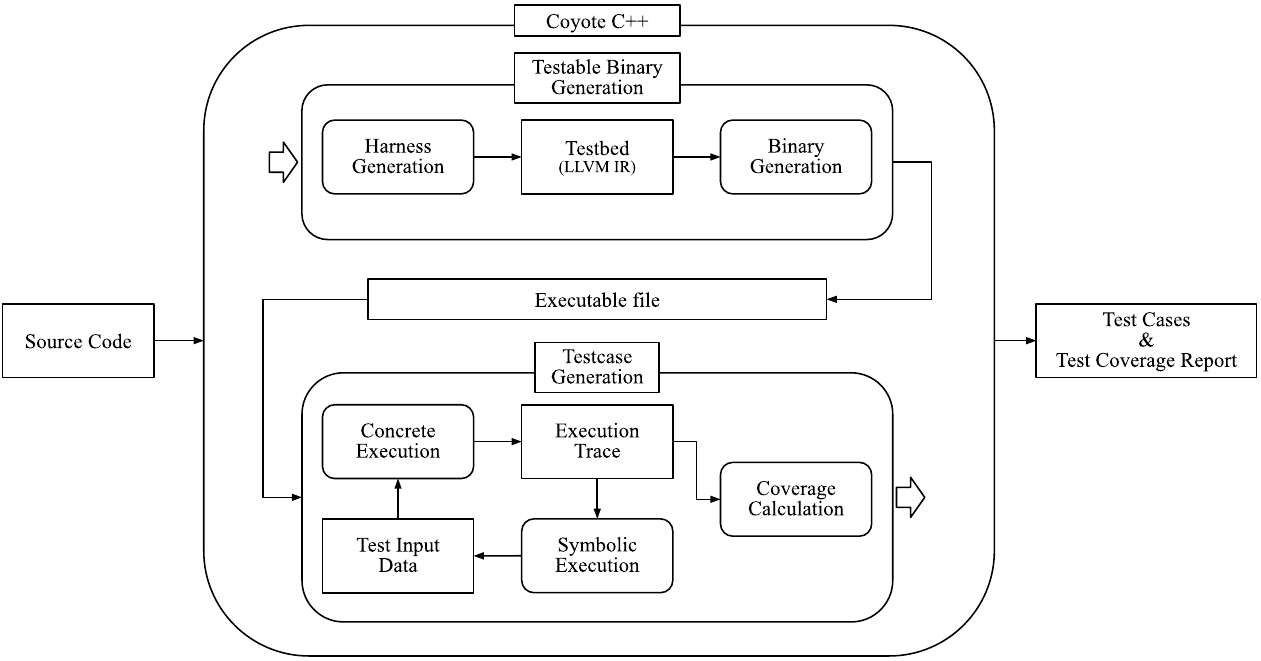}
    \caption{Overview of Coyote \CC.}
    \label{fig:overview}
\end{figure*}

In this chapter, we present an overview of Coyote \CC\ and discuss the core decisions that influenced its design. As shown in the diagram in Fig.~\ref{fig:overview}, the Coyote \CC\ tool is divided into two main parts. The first part builds executable test files based on harness generation, while the second part handles generating test cases through concolic execution.

In the first phase, Coyote \CC\ uses a harness generator module to automatically generate test stubs and test drivers for test execution and inserts instrumentation code for concolic execution. This instrumentation is performed on LLVM IR level. Next, the binary generation module compiles the created testbed to executable files used in the second part.

While running the executable test file in the second phase, the instrumentation code produces trace files containing information about the concrete program execution on the level of LLVM IR instructions. These trace files are then used to reconstruct their respective execution paths, and with this information symbolic execution is performed on the LLVM IR level to generate new test input data. When this concolic execution cycle is finished, the trace files are again used to compute the achieved coverage.

\subsection[Design Decisions of Coyote C++]{Design Decisions of Coyote \CC}

While implementing Coyote \CC, many important design decisions had to made. In the modules responsible for the testable binary generation, these decisions were generally made with the goal of enabling a wide range of transformations on intermediate code models while retaining a sufficiently strong connection between these models and the original source code. Most design decisions affecting the testcase generation phase were strongly influenced by the need to find a suitable tradeoff between the achieved code coverage and performance in terms of test time or resource consumption.

A fundamental design decision made in Coyote \CC\ is using LLVM IR as its symbolic execution target. This allows for more precision than doing source level symbolic execution while retaining more information about the original source code that would be lost when lowering even further to the assembly level. Also, using LLVM as a foundation for Coyote \CC\ allows for greater freedom in code transformations during harness generation, bypassing syntactic constraints present on the source code level.

We decided to implement offline testing by inserting instrumentation code into the LLVM IR code of the testbed during testable binary generation. The main reason for choosing offline testing over online testing is that it is more suitable for parallelization, which is essential for providing good testing performance. Additionally, offline testing is more advantageous from a memory management standpoint.

A key factor for achieving high code coverage is the search strategy that controls in which order the possible execution paths of a program are explored. During testcase generation, the test files are initially executed with all test inputs set to default values. The trace files generated from this are then analyzed using concolic execution techniques to create new test case inputs for visiting new paths. As our search strategy for exploring of candidate paths, we adopted a hybrid approach that combines CCS (Code Coverage Search) and DFS. CCS focuses on exploring code areas that have not been traversed yet, making it advantageous for quickly reaching high coverage. However, because CCS performs rather aggressive pruning on execution paths, it may produce unsatisfiable path conditions in certain situations. To make up for these issues, we also use the DFS strategy in addition to CCS. DFS is a search strategy that has the potential to cover code areas not covered by CCS, but it comes with the drawback of substantial time consumption and may even not terminate in worst-case scenarios.

Finally, a significant factor influencing the performance of concolic execution in \CC\ is the memory model. Similar to \textsc{Mayhem}, the approach implemented in Coyote \CC reads values from memory symbolically but writes values to concrete memory addresses. Utilizing symbolic reads in contrast to reading from concrete addresses leads to a more faithful representation of path constraints, thereby enhancing the potential for generating appropriate test cases. For write operations however, we chose to rely on concrete addresses because symbolic writes are prone to making the process of solving the path constraints overly expensive.

\begin{table*}[!htbp]
\caption{Results on Open-Source Projects}
\begin{center}
\begin{tabular}{|l|r|r|r|r|r|r|r|}
    \hline
    \multicolumn{5}{|c|}{\textbf{Project Info}} & \multicolumn{2}{c|}{\textbf{Coverage}} & \multicolumn{1}{c|}{\textbf{Test Time}} \\
    \cline{1-7}
    \multicolumn{1}{|c|}{\textbf{\textit{Name (C/\CC)}}} & \multicolumn{1}{c|}{\textbf{\textit{Files}}} & \multicolumn{1}{c|}{\textbf{\textit{Functions}}} & \multicolumn{1}{c|}{\textbf{\textit{Statements}}} & \multicolumn{1}{c|}{\textbf{\textit{Branches}}} & \multicolumn{1}{c|}{\textbf{\textit{Statement}}} & \multicolumn{1}{c|}{\textbf{\textit{Branch}}} & \multicolumn{1}{c|}{\textbf{[m]}} \\
    \hline
    nuklear (C)         &  39 &   609 &  9,284 &  4,309 & 93.7\% &  87.1\% & 55 \\
    libsodium (C)       &  94 &   887 &  8,003 &  1,651 & 96.5\% &  89.7\% &  6 \\
    mathc (C)           &   1 &   843 &  4,192 &    190 & 99.9\% & 100.0\% &  3 \\
    aubio (C)           &  53 &   520 &  5,916 &  1,797 & 95.7\% &  92.4\% & 14 \\
    s2n-tls (C)         & 175 & 1,621 & 16,734 & 15,512 & 86.7\% &  81.3\% & 68 \\
    yaml-cpp (\CC)      &  32 &   367 &  3,050 &  1,300 & 96.9\% &  95.5\% & 11 \\
    qnite (\CC)         &  48 &   637 &  4,294 &  1,035 & 95.2\% &  89.1\% & 37 \\
    json-voorhees (\CC) &  21 &   451 &  2,507 &    764 & 92.5\% &  88.7\% &  5 \\
    QPULib (\CC)        &  24 &   278 &  3,561 &  1,398 & 87.8\% &  83.8\% &  3 \\
    jsoncpp (\CC)       &   3 &   309 &  2,802 &  1,148 & 91.2\% &  86.3\% & 11 \\
    \hline
    \multicolumn{1}{|c|}{\textbf{Total}}      & 490 & 6,522 & 60,343 & 29,104 & 93.6\% &  89.4\% & 213 \\
    \hline
\end{tabular}
\label{tab:opensourcecov}
\end{center}

\bigskip

\caption{Coverage Results from Hyundai KEFICO}
\begin{center}
\begin{tabular}{|c|r|r|r|r|r|r|c|}
    \hline
    \multicolumn{5}{|c|}{\textbf{Project Info}} & \multicolumn{2}{c|}{\textbf{Coverage}} & \multirow{2}{*}{\textbf{Test Time}} \\
    \cline{1-7}
    \textbf{\textit{Name}} & \multicolumn{1}{c|}{\textbf{\textit{Files}}} & \multicolumn{1}{c|}{\textbf{\textit{Functions}}} & \multicolumn{1}{c|}{\textbf{\textit{Statements}}} & \multicolumn{1}{c|}{\textbf{\textit{Branches}}} & \multicolumn{1}{c|}{\textbf{\textit{Statement}}} & \multicolumn{1}{c|}{\textbf{\textit{Branch}}} & \\
    
    \hline
    Target A & 1,855 & 5,129 &  129,131 &  40,718 & 92.8\% & 86.8\% & \multirow{4}{*}{\textbf{N/A}} \\
    Target B & 83 & 1,774 &  11,828 &  3,078 & 97.4\% & 90.7\% & \\
    Target C & 69 & 375 &  6,526 &  2,339 & 85.5\% & 79.9\% & \\
    \cline{1-7}
    \textbf{Total} & 2,007 & 7,278 & 147,485 & 46,135 & 92.9\% & 86.7\% & \\
    \hline
\end{tabular}
\label{tab:kefico}
\end{center}
\end{table*}

\section{Experimental Results}

To showcase the performance of Coyote \CC, we present experimental results for a set of diverse open-source projects as well as several industrial software projects from one of our customers, Hyundai KEFICO. While our tool allows user to add test cases and write driver functions for achieving higher coverage, all experimental results were obtained through one-click automation without any user intervention.

\subsection{Experiment on Open-Source Projects}

For the first evaluation, we chose to reuse the test set curated by Shin and Yoo for a survey on white-box automated testing tools~\cite{kaistEval}, as it contains open-source projects written in C and \CC\ from a wide variety of application domains and was composed specifically for the evaluation of automated testing tools such as Coyote \CC. This survey also concluded that currently no other commercial tools truly support automated testing for \CC\ programs. Among open-source tools for \CC, CITRUS~\cite{Citrus} is no longer publicly available, and we were not able to successfully apply UTBot~\cite{Utbot} to the selected test projects due to its rather limited support for the \CC\ syntax. Thus, unfortunately there were no suitable candidates to compare Coyote \CC\ against in terms of coverage and test time.

Table \ref{tab:opensourcecov} shows the statement\footnote{As statements we consider only executable lines of code. In contrast to physical lines of code, this excludes e.g. whitespace, comments, and type declarations.} and branch coverage results achieved by Coyote \CC\ on the ten open-source projects in the test set as well as the time needed for conducting the automated test generation and execution for each project. Coyote \CC\ achieves statement coverages between 86.7\% (s2n-tls) and 99.9\% (mathc) as well as branch coverages between 81.3\% (s2n-tls) and 100\% (mathc). Summing up the number of overall covered lines/branches and dividing them by the total number of lines and branches in all ten projects yields a remarkable combined statement coverage of 92.5\% and branch coverage of 84.9\%.

The test times presented in table \ref{tab:opensourcecov} were attained from an Intel Core i7-13700 system with 64GB of RAM running Ubuntu 20.04. Overall, the test of all ten projects combined only took about three and a half hours, with individual testing times ranging between three minutes (mathc) and just above one hour (s2n-tls). That makes it more than six times faster than the test times reported in the previously mentioned study~\cite{kaistEval}, which we consider a significant improvement despite possible minor differences between test setups. Furthermore, with the exception of the qnite project, the testing speed on all projects surpasses our definition of practicality, with an overall testing speed of roughly 17,000 statements per hour.

\subsection{Results on Industry Projects}

Table \ref{tab:kefico} presents testing results produced by Coyote \CC\ on automotive control software projects from our customer Hyundai KEFICO, a member of Hyundai Motors Group. As details about these projects such as their actual names are strictly internal information, we will refer to them as target A, B and C.

The coverage results for these industrial projects are quite similar to the open-source projects, with an average statement coverage of 92.9\% and an average branch coverage of 86.7\%. At our customer, Coyote \CC\ is employed not in a controlled test environment but rather in a business setting on multiple machines with varying hardware specifications. Due to these circumstances and the fact that a subset of the test results were produced incrementally over a longer period of time, we presently do not have any meaningful test time measurements available to report for these projects.

While project C individually yields a slightly subpar coverage, our notion of practicality in terms of coverage achieved (statement coverage $>$90\%, branch coverage $>$80\%) is upheld both by projects A and B individually as well as all three projects combined. This again reinforces our claim that Coyote \CC\ is not simply a research prototype which only works on a limited set of specially curated programs but is rather a mature tool that can also handle more challenging industry software. Also, it should be noted that automated testing with such high coverage results for these three projects is only possible because Coyote \CC\ has explicit handling for some common code patterns in embedded software that would usually make automated testing difficult or plainly impossible, such as the usage of fixed memory addresses in code.

\section{Conclusion and Future Work}

In this paper, we presented Coyote \CC, an industry-grade automated testing tool based on concolic execution. After describing the general tool architecture, we discussed the core design decisions for our implementation of its concolic testing engine. Finally, we evaluated the performance of Coyote \CC\ in terms of achieved coverage and testing time on both a test set of diverse open-source projects and industry code from one of our corporate customers. We were able to demonstrate that Coyote \CC\ can achieve high statement/branch coverage of around 90\% or higher in a reasonable amount of time for software projects from a wide variety of application domains.

While Coyote \CC\ is already yielding promising results both on open-source projects and in real industry applications, it is our plan to continuously improve the tool both in terms of reliably achieving high coverage results and broadening its capabilities in the field of automated testing.

One goal for the near future is target testing for embedded software. Our tool currently performs host testing, meaning tests are not executed on the hardware that would run the program under test in a production environment, but rather on a separate computer, e.g., a test engineer's computer or a test server. Especially in the embedded domain however, the discrepancy between embedded hardware in the production environment and the consumer or server hardware in the testing environment may lead to inaccurate test results. Thus, we are planning to implement target testing support so that tests may be run directly on production hardware.

Approaching the goal of increasing automated test coverage from a different perspective, we also strive to provide users of our tool with feedback as to how they should change their code so that Coyote \CC\ will likely yield better coverage results for it. While we would like to give such guidance on the basis of code metrics, our initial investigations have shown that traditional code metrics such as cyclomatic complexity have little to no correlation with automated test coverage. Thus, we see the need for more thorough research involving the development of new code metrics that can serve as a better estimate for the coverage results produced by automated testing and Coyote \CC\ in particular.

\bibliographystyle{IEEEtran}
\bibliography{bibliography}

\end{document}